\documentclass[12pt]{article}

\usepackage{amsmath}
\usepackage{amsfonts}
\usepackage{amssymb}
\def\be{\begin{equation}}
\def\ee{\end{equation}}

\def\vp{\varphi}

\begin{document}


\begin{flushright} {\footnotesize HUTP-03/A036}  \end{flushright}
\vspace{5mm}
\vspace{0.5cm}
\begin{center}

\def\thefootnote{\fnsymbol{footnote}}

{\Large \bf On non-gaussianities in single-field inflation} \\[1cm]
{\large Paolo Creminelli}
\\[0.5cm]

{\small 
\textit{Jefferson Physical Laboratory, \\
Harvard University, Cambridge, MA 02138, USA}}

\end{center}

\vspace{.8cm}

\hrule \vspace{0.3cm} 
{\small  \noindent \textbf{Abstract} \\[0.3cm]
\noindent
We study the impact of higher dimension operators in the inflaton Lagrangian on the non-gaussianity of 
the scalar spectrum. These terms can strongly enhance the effect without spoiling slow-roll, though it is 
difficult to exceed $f_{\rm NL} \sim 1$, because the scale which suppresses the operators cannot 
be too low, if we want the effective field theory description to make sense. In particular we explicitly
calculate the 3-point function given by an higher derivative interaction of the form $(\nabla\phi)^4$,
which is expected to give the most important contribution. The angular dependence of the result turns out 
to be quite different from the minimal case without higher dimension operators.}

\vspace{0.5cm}  \hrule

\def\thefootnote{\arabic{footnote}}
\setcounter{footnote}{0}


\section{Introduction}
In most models, inflation takes place at energies many orders of magnitudes higher than those 
accessible with accelerators. It is therefore mandatory to explore all what we can learn about this
high energy regime from the signatures left by inflation in the present Universe. In the effective
field theory framework the effects of short distance physics can always be encoded in higher dimension 
operators for the inflaton Lagrangian\footnote{Another logical possibility is that unknown high energy 
physics modifies the vacuum state of the inflaton in the approximately de-Sitter background of inflation 
\cite{Brandenberger:2002hs}. We will not consider this possibility here, assuming that the inflaton is 
in the usual Bunch-Davies vacuum.} \cite{Kaloper:2002uj}. The obvious problem is how to experimentally probe
the presence of these operators. Despite the successes of inflationary cosmology in fact, there are very few 
observables which can be used to constrain the inflaton dynamics. 

References \cite{Kaloper:2002uj} and \cite{Shiu:2002kg} studied the possibility that higher
dimension operators can change the prediction for the tilt of the gravitational wave spectrum. This 
quantity is fixed, in single field models, once the ratio between scalar and tensor modes is known,
through the so-called 'consistency relation'. The problem with this kind of observable is that the
detection of gravitational waves is possible only if the scale of inflation is close to the presently
maximum allowed value, while most models predict a totally negligible production of tensor modes. Even
if we restrict to models with a sensible production of gravity waves, the tilt of this spectrum remains
almost impossible to probe. Order one deviation from the consistency relation could be measurable
with dedicated polarization experiments as CMBpol, only if the tensor to scalar ratio is very big:
$T/S \gtrsim 0.1$ \cite{Knox:2003jw}. 

The purpose of this note is to indicate a potentially better test of higher dimension operators in
the inflaton Lagrangian, which is not related to the elusive tensor modes: non-gaussianities in the 
scalar perturbations. The basic reason why this could be a good smoking gun of high energy effects
in inflation is that it is quite suppressed in the standard framework. This follows from the fact that the 
inflaton must be {\em very} weakly coupled to have a slow-roll phase, so that its fluctuations are
very close to those of a free field, {\em i.e.} gaussian. 

Consider for example the effect of the cubic self-interaction of the inflaton $\phi$, given by 
$V^{(3)}(\phi)$, where V is the inflaton potential. The amount of non-gaussianity is parametrized 
in this case by $V^{(3)}(\phi)/(2 \pi H)$, with $H$ the Hubble constant during inflation. In the equation of 
motion for the inflaton fluctuations, this ratio gives the relative importance of the non-linear term 
with respect to the free field terms at horizon crossing. The slow-roll conditions impose a strong 
bound on the derivatives of the potential; in particular the constraint on the third derivative is
\begin{equation}
\label{eq:slowroll}
M_P^4 \frac{V' \; V^{(3)}}{V^2} \ll 1 \quad \Rightarrow \quad \frac{V^{(3)}}{2 \pi H} \ll 
\frac3{2 \sqrt2 \pi} \frac1{\sqrt{\epsilon}} \frac{H}{M_P} = 3 P_{\cal{R}}^{1/2}\quad 
\qquad {\rm with} \quad \epsilon \equiv \frac12 M_P^2 \left(\frac{V'}{V}\right)^2 \;.
\end{equation}
The limit is quite strong and it is completely model independent, because $P_{\cal{R}}$ is the measured
curvature power spectrum: $3 P_{\cal{R}}^{1/2} \simeq 1.4 \cdot 10^{-4}$. 
We therefore see that the existence of a slow-roll phase requires
that the inflaton fluctuations are very close to gaussian. 

The full calculation of the 3-point function has been carried out quite recently in 
\cite{Maldacena:2002vr} and \cite{Acquaviva:2002ud}. It turns out that the strong constraints we described 
on the inflaton potential make the inflaton coupling to gravity the leading source of non-gaussianity.
The result is suppressed with respect to (\ref{eq:slowroll}) by a combination of the slow-roll
parameters. Parametrically the level of non-gaussianity is thus predicted to be
\begin{equation}
\label{eq:Malda}
{\cal{O}}(\epsilon, \eta) \cdot \frac1{2 \pi}\frac1{\sqrt{\epsilon}} \frac{H}{M_P} \lesssim 3 \cdot 10^{-6}  
\quad \qquad {\rm with} \quad \eta \equiv  M_P^2 \frac{V''}{V} \;.
\end{equation}
Comparing this prediction with the current limit coming from the WMAP satellite \cite{Komatsu:2003fd}, 
which is of order $5 \cdot 10^{-3}$, we see that the level of non-gaussianity in conventional inflationary
models is quite low and certainly it will not be measured in near-future experiments. This very
low prediction is clearly good to test deviations from the simplest models.  

A possibility to get sensible non-gaussianities in to have an additional light field 
({\em i.e.}~with mass smaller than $H$) during inflation. This new degree of freedom is not constrained
from the slow-roll requirements and if its fluctuations are somehow converted into 
metric perturbations, these can be much less gaussian than in the usual scenario. This possibility is 
realized in multiple-field models of inflation \cite{Bartolo:2001cw,Bernardeau:2002jy}, in the curvaton 
scenario \cite{Lyth:2002my} and in the recently proposed mechanism with curvature perturbations created 
by fluctuations of the decay rate of the inflaton \cite{Dvali:2003em,Zaldarriaga:2003my}.

We will follow a more minimal possibility here. We want to see if the addition of higher-dimension
operators in otherwise conventional models can give an increase of the result (\ref{eq:Malda}).  
Before concentrating on an explicit calculation in section \ref{sec:calc}, we discuss which operators 
we are allowed to add in an effective field theory description of inflation. 

\section{Effective field theory during inflation}

During inflation there are two energy scales of interest. One is related to the classical time 
evolution of the inflaton and it is given by $\dot\phi^{1/2}$. The other one is typical of the quantum
fluctuations of the inflaton around this classical background and it is given by the Hubble scale
$H$. The separation of these two energy scales is fixed by the normalization of the scalar perturbations
originally measured by COBE: $H/\dot\phi^{1/2} \sim 10^{-2}$. As we want to encode unknown 
short distance physics in an effective field theory description, it is natural to ask which is the
typical scale of this theory or equivalently which heavy states we are allowed to integrate out. 

In principle, as we are interested in the study of the inflaton perturbations, which are characterized
by $H$, we could be tempted to write down a generic Lagrangian for the fluctuations, 
with higher dimension operators suppressed by a scale which is only required to be bigger than $H$. 
This is for example the attitude taken in \cite{Kaloper:2002uj}. 
This procedure is too arbitrary. Many constraints on the Lagrangian come from the requirement of 
having a slowly rolling background or, in other words, it is not easy to imagine that any effective 
operator for the inflaton perturbations can be obtained by a UV completion compatible with slow-roll.
Moreover if we took this point of view, there would be no reason to expect nearly gaussian 
perturbations at all, because we could write down a generic self-interaction for the inflaton, while
we saw that the constraints come from the required flatness of the potential in order to have an
inflating phase.

Our approach will be to consider a generic effective Lagrangian, which encodes all the heavy physics 
above the higher scale $\dot\phi^{1/2}$. In principle we are allowed to write down all operators 
we want, with the only constraint that an inflating phase is possible. Every choice will modify the 
background evolution and the fluctuations at the same time. 

The question now is which operators can give big non-gaussianities without spoiling the slow-roll of 
the inflaton. References \cite{Maldacena:2002vr,Acquaviva:2002ud} conclude that for a generic potential
we get the estimate (\ref{eq:Malda}). We are therefore led to consider terms with derivatives. It is
not difficult to conclude that operators with derivatives acting {\em on each} field $\phi$, {\em i.e.}
compatible with a shift symmetry $\phi \rightarrow \phi + c$, are the best candidates because they
do not interfere with the slow-roll evolution of the inflaton.
In particular one could impose an approximate shift symmetry on the inflaton, so that {\em only}
these terms are allowed in the limit of exact symmetry. Obviously the symmetry cannot be 
exact otherwise the inflaton would be a Goldstone boson with a perfectly flat 
potential; some tilt must be introduced through an explicit breaking, making the inflaton a 
pseudo Goldstone boson\footnote{The idea of the inflaton as a pseudo Goldstone boson was
introduced a long time ago in \cite{Freese:1990rb} and recently reinterpreted 
in the context of higher-dimensional (or higher-dimensionally inspired) ideas 
\cite{Arkani-Hamed:2003wu}.}. 

It is therefore interesting to study the set of all operators compatible with the exact shift
symmetry of the inflaton\footnote{The effects of higher derivative interactions on the consistency relation 
have been studied in \cite{Kaloper:2002uj,Shiu:2002kg}.}, suppressed by the appropriate power of the scale 
of new physics $M$. These derivative interactions could arise, for example, from a Higgs UV completion,
where the shift symmetry is derived from a linearly realized, spontaneously broken U(1) invariance, or
they could be the consequence of strong dynamics at the scale $M$. We stress that these
operators, modifying just the kinetic part of the inflaton equation of motion are not dangerous for 
the slow-roll and we are allowed to take the scale $M$ to be as low
as $\dot\phi^{1/2}$: when this limit is reached all the higher-dimension operators are equally important
for the classical evolution of the inflaton and the effective field theory description ceases to be 
useful\footnote{We stress that it would be possible to add specific operators with many derivatives 
(as $\Box\phi \cdot (\nabla\phi)^2$ for example) suppressed by a lower mass scale to get big non-gaussianities 
without spoiling slow-roll.
On the other hand, without an explicit UV completion, we are forced to take all operators compatible 
with the given symmetries, suppressed by the same scale of new physics $M$ and to remain in the regime 
in which this effective field theory description makes sense. It is conceivable, but unlikely, that a 
specific UV completion gives a much bigger non-gaussianity than what obtained here.}.

Among all the operators compatible with the shift symmetry, the most important are those of the
form $(\nabla\phi)^{2 n}$. Operators with more derivatives acting on one field 
(as $\Box\phi \cdot (\nabla\phi)^2$) are less relevant because each derivative gives a suppression 
of order $H/M$, both in studying the evolution of the classical background and for the 
fluctuations around it. In particular the lowest dimension operator will be $(\nabla\phi)^4/M^4$. 

What is the size of the 3-point function we expect adding this higher-dimension operator?
One of the four legs will be evaluated on the classical background $\dot\phi \simeq V'/H$, 
so that we obtain an interaction among three fields with derivatives acting on them; each derivative
gives a factor of $H$ at horizon crossing. The corrections to the free theory, which control 
the deviation from exact gaussianity, are thus given by
\begin{equation}
\label{eq:estimate}
\frac{V'}{H} \cdot \frac{H^3}{M^4} \cdot \frac1{2 \pi H}  \lesssim \frac1{2 \pi}\frac{V' H}{\epsilon \;V} 
\simeq \frac1{2 \pi} \frac1{\sqrt{\epsilon}} \frac{H}{M_P} \;, 
\end{equation}
where the upper limit comes from setting $M^4 = \dot\phi^2 \simeq \epsilon \; V$ (\footnote{In the same 
limit we obtain order one deviations from the consistency relation \cite{Shiu:2002kg}.}). 
As we discussed, in this limit all the higher terms become important and the effective field theory 
description breaks down. 

We see that the effect can be much bigger than the result obtained in the minimal case 
(\ref{eq:Malda}), which is further suppressed by a slow-roll parameter. The amount of non-gaussianity
is still small ($\lesssim 5 \cdot 10^{-5}$) and, as we will discuss, it is close to the limit of what 
is measurable. To make more precise statements about the size of the effect and to study the
angular dependence of the 3-point function we now turn to the explicit calculation.

\section{\label{sec:calc}Calculation of the 3-point function}
The action we will consider is
\begin{equation}
\label{eq:action}
S = \int d^4 x \sqrt{-g} \left[ \frac12 M_P^2 R - \frac12 (\nabla\phi)^2 - V(\phi) + 
\frac1{8 M^4} (\nabla\phi)^2 (\nabla\phi)^2 \ldots \right] \;,
\end{equation}
where the dots stand for additional higher dimension operators compatible with the shift 
symmetry $\phi \rightarrow \phi + c$.
The background solution for the metric is of the form
\begin{equation}
\label{eq:metric}
ds^2 = -dt^2 + e^{2 H t} dx^i dx_i \;,
\end{equation}
with $H$ a slow-varying function of time ($\dot H/ H^2 \ll 1$).
From the action (\ref{eq:action}) it is easy to get the equations describing the evolution 
of the scalar $\phi$ and $H$ \cite{Armendariz-Picon:1999rj,Shiu:2002kg}
\begin{eqnarray}
\label{eq:H2} & & H^2 = \frac13 M_P^{-2} \left[V(\phi) + \frac12 \dot\phi^2 + \frac38 
\frac{\dot\phi^4}{M^4}\right] \\
\label{eq:phiev} & & \left[1 + \frac32 \frac{\dot\phi^2}{M^4}\right] \ddot \phi + 3 H \dot\phi
\left[1 + \frac12 \frac{\dot\phi^2}{M^4} \right] + V'(\phi) = 0 \;.
\end{eqnarray}
As we are interested in the calculation of the 3-point function at first order in $\dot\phi^2/M^4$,
we will often neglect the $1/M^4$ corrections to eqs.~(\ref{eq:H2}) and (\ref{eq:phiev}). In this 
limit the background equations reduce to the usual ones describing single-field inflation.

To study the perturbations around the background solution we follow \cite{Maldacena:2002vr}, using 
the ADM metric 
\begin{equation}
\label{eq:ADM}
ds^2 = -N^2 dt^2 + h_{ij} (dx^i + N^i dt)(dx^j + N^j dt) \;.
\end{equation}
In this formalism the dynamical degrees of freedom are $h_{ij}$ and the scalar $\phi$, while $N$
and $N^i$ have no time derivatives in the action, so that they behave as Lagrange multipliers.

We now have to fix the gauge to second order in the fluctuations. As discussed in \cite{Maldacena:2002vr},
there are two useful choices. The first one is given by\footnote{As we are not interested
in the tensor modes, we will not specify the parametrization of $\hat h_{ij}$. 
See \cite{Maldacena:2002vr}.}
\begin{equation}
\label{eq:zetagauge}
\delta\phi = 0 \;, \qquad h_{ij} = e^{2 H t + 2\zeta} \hat h_{ij} \;, \qquad \det \hat h =0 \;.
\end{equation}
The variable $\zeta$ is the non-linear generalization of the one introduced by Bardeen,
Steinhardt and Turner \cite{Bardeen:qw}. In references \cite{Salopek:1990jq} and 
\cite{Maldacena:2002vr}, it is proven that $\zeta$ is constant outside the horizon. For wavelengths
much bigger than the Hubble radius, we can neglect all gradient terms, so that different parts of the 
Universe follow the same evolution, with a certain shift in time parametrized by $\zeta$; this explains 
intuitively why this variable stays constant outside the horizon. Note that the same
result holds true in our case, after the addition of the higher derivative terms: these
interactions always contain derivatives acting on the fluctuations around the background, so that
they are irrelevant outside the horizon.

The other useful choice of gauge is given by
\begin{equation}
\label{eq:phigauge}
\phi = \phi(t) + \vp(t,x) \;, \qquad h_{ij} = e^{2 H t} \hat h_{ij} \;, \qquad \det \hat h =0 \;.
\end{equation} 
This gauge has an important property that will be useful for our calculation. At leading order in the 
slow-roll parameters the action for $\vp$ can be calculated from (\ref{eq:action}), without considering 
the gravitational contributions contained in $\sqrt{-g} R$ \cite{Maldacena:2002vr}. 
This is quite intuitive, because in the limit of a completely flat potential $\vp \neq 0$ does not
give any spacetime curvature. 

In this gauge the three point function $\langle\vp \vp \vp \rangle$ can therefore be calculated at
leading order in the slow-roll parameters from the interaction Lagrangian
\begin{equation}
\label{eq:Lint}
{\cal{L}_{\rm int}} = - \frac{e^{3 H t}}{2 M^4} (\dot\phi \dot\vp) (-\dot\vp^2 + 
e^{-2 H t} (\partial_i \vp)^2) \;,
\end{equation}
where one of the four legs of the $(\nabla\phi)^4$ interaction in (\ref{eq:action}) is
evaluated on the time-dependent background $\dot\phi$. As discussed we can disregard, at leading order,
the $1/M^4$ corrections to the background in eqs.~(\ref{eq:H2}) and (\ref{eq:phiev}). 

Before calculating $\langle\vp \vp \vp \rangle$, we have to discuss an important point. We are
interested in the $\langle\zeta \zeta \zeta \rangle$ correlator: this remains constant outside
the horizon and reappears as a non-gaussian contribution to the perturbations we observe today. Once
we cross the horizon we have therefore to go from the $\vp$ variable to $\zeta$ and 
this must be done at second order in the pertubations: a second order contribution could
give an additional piece $\langle\zeta \zeta \zeta \rangle \propto \langle \vp \vp \rangle 
\langle \vp \vp \rangle$. This is what happens in the calculation \cite{Maldacena:2002vr}
without higher derivative terms; the relation between $\zeta$ and $\vp$ is schematically given 
by 
\begin{equation}
\label{eq:etavp}
\zeta = - \frac{H}{\dot\phi} \vp + {\cal{O}}(\epsilon,\eta) \left(\frac{H}{\dot\phi} 
\vp\right)^2 \;.
\end{equation} 
The non-linear term is proportional to the slow-roll parameters and it exactly cancels the
$\vp$ evolution outside the horizon to give a time-independent $\zeta$. We will neglect
this non-linear term in our calculation as it is sub-leading in the slow-roll expansion. 
Moreover the relation $\zeta(\vp)$ is not modified outside the horizon by the addition of the 
higher dimension operator $(\nabla \phi)^4/M^4$: as we discussed, this term contains derivatives
acting on the fluctuations and it is therefore exponentially irrelevant when the wavelength exceeds 
the Hubble radius. So we can take just the linear term in (\ref{eq:etavp}) to convert the result
in the $\zeta$ variables after horizon crossing.

We now turn to the explicit calculation\footnote{For all the details about this kind of calculation
we refer the reader to the very clear discussion of \cite{Maldacena:2002vr}, whose notation we have
mostly followed.} of $\langle\vp \vp \vp \rangle$. 
At leading order in slow-roll we can take $H$ constant and we can write the contribution to 
$\langle\vp \vp \vp \rangle$ from the interaction (\ref{eq:Lint}) as an integral over the conformal
time $\eta$:
\begin{align}
\label{eq:phiintegral}
\langle\vp_{\vec k_1} \vp_{\vec k_2}\vp_{\vec k_3}\rangle & = (2\pi)^3 \delta^3\Big(\sum_i \vec k_i\Big) 
\frac1{\prod(2 k_i^3)} \frac{\dot\phi H^5}{2 M^4} \cdot \\ \nonumber & \cdot i \int_{-\infty}^0 d\eta 
\big(-k_1^2 k_2^2 k_3^2 \;\eta^2 - (\vec k_1 \cdot \vec k_2) k_3^2 (1 - i k_1 \eta)(1 - i k_2 \eta)\big) 
e^{i k_t \eta} + {\rm perm.} + c.c.
\end{align}
where $k_i$ are the moduli of the wavevectors and $k_t = k_1 + k_2 + k_3$. The integration
is done with an infinitesimal contribution in imaginary time ($\eta \rightarrow \eta + i
\epsilon |\eta| $); this prescription projects the initial state in the usual Bunch-Davies
vacuum at past infinity. 

Note that the perturbations $\vp$ are strongly coupled well inside the horizon when their typical 
wavelength is of order $M$: all the higher dimension operators become important. Does this imply that 
our results depend on the UV completion of the theory? Obviously no. 
Whatever the UV completion might be, the evolution of the background is very slow with respect with 
the typical scale of this unknown physics because $M \gg H$: we therefore expect that, following the 
slow expansion, the system is always in its vacuum state. Although we do not know how this state looks 
like for very short wavelengths, when the description in terms of a scalar $\vp$ breaks down, we know 
that below the scale $M$ the system is in the Bunch-Davies vacuum of the weakly interacting $\vp$ field.
The same situation is at the origin of the so called 'trans-Planckian problem': the description of the
inflaton as a weakly interacting field ceases to make sense at the Planck scale. Even if the 
details of quantum gravity are unknown, it is likely that all the predictions are independent 
of this regime, as in our simple example.

Returning to the calculation, after the transformation to the $\zeta$ variable through 
$\zeta = - \frac{H}{\dot\phi} \vp$ we get the final result performing the integral in (\ref{eq:phiintegral})
\begin{align}
\label{eq:result}
\langle\zeta_{\vec k_1} \zeta_{\vec k_2}\zeta_{\vec k_3}\rangle & = (2\pi)^3 \delta^3\Big(\sum_i \vec k_i\Big) 
\frac1{\prod(2 k_i^3)} \frac{H^8}{\dot\phi^2 M^4} \cdot \\ \nonumber & \cdot
\frac1{k_t^2} \Big(\sum_i k_i^5 + \sum_{i \neq j} (2 k_i^4 k_j
- 3 k_i^3 k_j^2) + \sum_{i \neq j \neq l} (k_i^3 k_j k_l -4 k_i^2 k_j^2 k_l) \Big) \;.
\end{align}

As one could have expected the angular dependence given by the second line of (\ref{eq:result}) is 
completely different both from the result of the minimal case \cite{Maldacena:2002vr} and from that
assumed in the analysis of the data \cite{Komatsu:2003fd} which follows from the assumption that the 
non-gaussianities come from a field redefinition 
\begin{equation}
\label{eq:fieldred}
\zeta = \zeta_g -\frac35 f_{\rm NL} \zeta_g^2 \;,
\end{equation}
with $\zeta_g$ gaussian\footnote{This pattern of non-gaussianity which is local in real space is characteristic
of models in which the non-linearities develop outside the horizon. This happens if the perturbations are created 
by fluctuations in the reheating efficiency \cite{Dvali:2003em,Zaldarriaga:2003my} and in the curvaton scenario 
\cite{Lyth:2002my}.}. 
The most relevant difference is the behavior in the limit in which one of the 
wavevectors goes to zero. In this limit one of the fluctuations has a very big wavelength, it exits the
horizon and freezes much before the other two and acts as a sort of background. What do we expect in this 
limit? Let us take $k_3$ very small and consider the case in which a spatial derivative $\partial_i \vp$
in the interaction Lagrangian (\ref{eq:Lint}) is acting on $\vp_3$. The 2-point function 
$\vp_1 \vp_2$ depends on the position on the background wave and it is proportional 
to $\partial_i \vp_3$ at linear order. This variation of the 2-point function along the $\vp_3$ wave
is averaged to zero in calculating the 3-point function 
$\langle\vp_{\vec k_1} \vp_{\vec k_2}\vp_{\vec k_3}\rangle$, because the spatial average 
$\langle\vp_3 \partial_i\vp_3\rangle$ vanishes. So we are forced to go to the second order and we therefore
expect that the second line of (\ref{eq:result}) behaves as $k_3^2$ and not as $k_3$ in the limit 
$k_3 \rightarrow 0$. Note that this result holds also when a time derivative acts in (\ref{eq:Lint}) 
on the long wavelength mode, because the scalar wavefunction in de-Sitter space goes to a constant 
outside the horizon as the square of the physical wavevector $k_3^2/(a H)^2$.
Although it is not evident, it is easy to check that the second line of (\ref{eq:result}) does indeed go 
to zero as $k_3^2$ when $k_3 \rightarrow 0$. This is a radical difference from the behaviour studied in
\cite{Maldacena:2002vr} and \cite{Komatsu:2003fd}: they both go to a constant in this limit. The difference is a
consequence of the derivative interaction $(\nabla \phi)^4$, which favors the correlation among modes 
of comparable wavelength.

The different angular behaviour will probably change the limits coming from data analysis. It is anyway
interesting to calculate the ``effective'' $f_{\rm NL}$ in (\ref{eq:fieldred}) in the case of an 
equilateral triangle: $k_1 = k_2 = k_3$. It is easy to get
\begin{equation}
\label{eq:f_NLeq}
f_{\rm NL}^{\rm equil.} = \frac{35}{108} \frac{\dot\phi^2}{M^4} \;. 
\end{equation}
The result is not far from the naive estimate we did in the introduction and it should be compared with
what is obtained in absence of higher derivative terms: in the same configuration 
ref.s \cite{Maldacena:2002vr} and \cite{Acquaviva:2002ud} obtain 
\begin{equation}
\label{eq:f_NLst}
f_{\rm NL}^{\rm equil.} = \frac56 \big(\eta - \frac{23}6 \epsilon\big) \;.
\end{equation}
Comparing (\ref{eq:f_NLeq}) and (\ref{eq:f_NLst}) we see that for sufficiently low mass scale $M$ the
new contribution dominates over the usual one, which is suppressed by the slow-roll parameters. As
we discussed it is not possible to take $M$ arbitrarily small, because at a certain point the expansion
$\dot\phi/M^2$ ceases to make sense and we should consider all the higher dimension operators in the
action (\ref{eq:action}). We can estimate the lower limit on $M$ assuming that there is some strong
coupling regime around $M$ and using naive dimensional analysis to estimate the size of the 
operators. In this case we expect a Lagrangian given by
\begin{equation}
\label{eq:NDA}
\frac{\Lambda^4}{16 \pi^2} \left(\frac12 \frac1{\Lambda^2}(\nabla\tilde\phi)^2 + 
\frac18 \frac1{\Lambda^4} (\nabla\tilde\phi)^4 + \frac1{48} \frac1{\Lambda^6} (\nabla\tilde\phi)^6 \cdots 
\right) \;.
\end{equation}
In front of each of operator unknown order one coefficients should be understood. Going from the 
dimensionless scalar $\tilde\phi$ to the canonical normalization we obtain that the expansion makes sense 
for $\dot\phi/M^2 \lesssim 1$. The same conclusion holds if we take a UV completion where $\phi$
represents the Goldstone boson of a spontaneously broken U(1) symmetry. 

Even if we could have some order one enhancement, {\it e.g.} if the $(\nabla \phi)^4$ operator is 
somewhat bigger than its natural size, we conclude that a reasonable limit on non-gaussianities in
single-field models is
\begin{equation}
\label{eq:limit}
f_{\rm NL} \lesssim 1 \;.
\end{equation}

This limit is roughly what is expected to be measurable with the forecoming CMBR experiments. 
With the new data WMAP will reach a sensitivity $|f_{\rm NL}| \lesssim 20$, while the Planck 
satellite will improve the limit to $|f_{\rm NL}| \lesssim 5$ \cite{Komatsu:2002db}. Note that the latter result
is not far from what can be measured {\em in principle} in a CMBR experiment. The number of independent
spots in the sky is limited by the diffusion length of photons before recombination and Planck
sensitivity will not be far from this limit. 
Large scale structure surveys are not so severely limited by cosmic variance, but it is seems difficult that they 
can reach limits on primordial non-gaussianities stronger than those coming from CMBR experiments
\cite{Verde:1999ij}. The limit (\ref{eq:limit}) is therefore at the threshold of detectability. 

We have to stress that many additional effects are expected to give $f_{\rm NL} \sim 1$, even
starting from a gaussian initial perturbation. All the non-linearities in the evolution from
the imprinting during inflaton to the detection of temperature fluctuations will give $f_{\rm NL} \sim 1$,
as it is clear from the definition (\ref{eq:fieldred}). These effects should be under control 
before a clear signal of primordial non-gaussianity can be found. The characteristic angular behaviour 
(\ref{eq:result}) can be a powerful discriminator against all the other effects. As the 3-point
function contains in principle much more information than the single parameter $f_{\rm NL}$, it would be 
very interesting if the data analysis were carried out considering different angular patterns 
\cite{Gaztanaga:2003hs}.

\section{Conclusions}

Due to the strong suppression of non-gaussianities in minimal models of inflation, a deviation from
this prediction is probably the most efficient method to detect the effect of short distance physics during
inflation, which can change the inflaton dynamics through higher dimension operators. 

We studied the effect of operators compatible with a shift symmetry of the inflaton as the most 
promising set of higher dimensional terms. The explicit calculation of the three point function 
(\ref{eq:result}) shows a particular angular dependence as a consequence of the derivative interactions. 
Although the size of the operators can only be calculated once the UV completion is known, we conclude that
$f_{\rm NL} \sim 1$ represents the upper limit of this effect, before the effective field theory  
description breaks down. This limit is close to what will be measured by the Planck experiment. 

This result can be considered in some sense an upper limit for single field models. A detection of a much 
bigger non-gaussianity, $f_{NL} \gg 1$, would point towards models with additional light degrees of freedom 
during inflation.

\section*{Acknowledgments}
 It is a pleasure to thank Nima Arkani-Hamed, Hsin-Chia Cheng, Nemanja Kaloper, Shinji Mukohyama, 
 Lisa Randall and Matias Zaldarriaga for many useful discussions.


\end{document}